\documentclass[12pt]{article}

\newcommand{\x}{arXiv:}
\newcommand{\m}{\mathrm}
\newcommand{\be}{\begin{equation}}
\newcommand{\ee}{\end{equation}}
\newcommand{\ba}{\begin{eqnarray}}
\newcommand{\ea}{\end{eqnarray}}

\usepackage{graphicx}
\usepackage{amssymb}
\usepackage{amsmath}
\usepackage[T1]{fontenc} %for \boldsymbol
\usepackage[ansinew]{inputenc} %for \boldsymbol
\usepackage[nosort]{cite}
\newcommand{\inbar}{\vrule height1.57ex width.4pt depth0pt}
\newcommand{\SW}{\relax{\hbox{$\ \inbar\kern-.285em{\rm S}$}}}

\begin{document}
\thispagestyle{empty}
\begin{center}

\null \vskip-1truecm \vskip2truecm

{\Large{\bf \textsf{Planar Black Holes as a Route to Understanding the Weak Gravity Conjecture}}}

{\large{\bf \textsf{}}}

{\large{\bf \textsf{}}}

\vskip1truecm

{\large \textsf{Brett McInnes}}

\vskip1truecm

\textsf{\\  National
  University of Singapore}

\textsf{email: matmcinn@nus.edu.sg}\\

\end{center}
\vskip1truecm \centerline{\textsf{ABSTRACT}} \baselineskip=15pt
\medskip

One version of the Weak Gravity Conjecture requires that it should be possible for an extremal black hole to emit a smaller black hole: that is, the original black hole bifurcates. For asymptotically flat and asymptotically AdS Reissner-Nordstr\"{o}m black holes with spherical event horizons, such a bifurcation reduces the total classical entropy of the system, and so it is apparently forbidden by the second law of thermodynamics. It may well be possible to remedy this by taking other (for example, quantum-gravitational) effects into account, but it is difficult to understand this in a quantitative way. In the case of asymptotically AdS Reissner-Nordstr\"{o}m black holes with \emph{planar} event horizons, however, one can show that bifurcations are definitely compatible with the second law. (Naked singularities, generated by the bifurcation, may play an important role here.) Furthermore, in this case one can exhibit a detailed mechanism explaining precisely why planar black holes must indeed be unstable (through emission of branes) when they are sufficiently close to extremality. Thus planar black holes can improve our understanding of the WGC.

\newpage

\newpage

\addtocounter{section}{1}
\section* {\large{\textsf{1. Black Hole Bifurcations }}}
A theorem in classical General Relativity (see \cite{kn:wald}, Theorem 12.2.1) is usually interpreted as stating that, provided that a spacetime is ``predictable'' (in the sense that no information enters from, in particular, naked singularities\footnote{The precise condition is that of \emph{strong asymptotic predictability}: see \cite{kn:wald}, page 299. For the interpretation of this condition as meaning that naked singularities are absent, see page 301 of the same reference.}) then a black hole in that spacetime cannot bifurcate as time passes. To put it the other way around: a black hole which bifurcates as time passes will inevitably generate a naked singularity, that is, some kind of (unspecified) violation of Cosmic Censorship. Let us call this the ``Bifurcation Theorem''.

The theorem means that there is a price to be paid if a theory requires black holes to bifurcate. In the past, this price, the formation of a naked singularity, would normally have been judged too high; but recent developments may lead us to reconsider this judgement.

In particular, the Bifurcation Theorem is directly relevant to the \emph{Weak Gravity Conjecture} or WGC. This is based on the claim that extremal black holes should be unstable, despite having a zero Hawking temperature. In the first instance \cite{kn:motl} it was assumed that this involves the emission of \emph{particles} (or, as we will argue, \emph{branes}) with charge or angular momentum, depending on the reason for which the black hole is extremal. However, it has also been argued that the emitted object might itself be a charged \cite{kn:kats,kn:NAH} or rotating \cite{kn:rem2,kn:aalsma,kn:104,kn:105,kn:aalsma2,kn:canorui} \emph{black hole}. In this second case the WGC involves black hole bifurcations, and consequently it demands the existence of naked singularities (or of whatever might replace them in a full quantum-gravitational treatment\footnote{Instead of constantly repeating this phrase, we take it as understood that the expression, ``naked singularity'' is to be interpreted in this manner.}).

We need to stress at this point that the Bifurcation Theorem is very different from the familiar Singularity Theorems: \emph{it does not rely on any energy condition} (see the proof given in \cite{kn:wald}). It is a relatively straightforward consequence of very basic aspects of spacetime causal structure, such as the fact that the interiors of lightcones are connected. This is important here, because an inherently quantum effect like the WGC might well involve violations of classical energy conditions. We see that, even if that is so, the Bifurcation Theorem still applies.

The possibility that naked singularities might arise when black holes evolve has of course been investigated extensively: see for example \cite{kn:ulrich} for the current status in the asymptotically flat case. Whether and under what circumstances naked singularities can arise in the asymptotically AdS case \cite{kn:toby} is a complex question\footnote{The conditions of the Bifurcation Theorem are, strictly speaking, not satisfied in the asymptotically AdS case or indeed in any spacetime which is not asymptotically flat, such as a closed FRW cosmology. However, it can be argued that it holds for isolated systems in such cases (see \cite{kn:wald}, page 301), and we will assume this here.} which, in fact, actually involves the WGC itself \cite{kn:weak,kn:horsant}. Recently however it was argued convincingly \cite{kn:roberto} that naked singularities can occur, even when asymptotically AdS black holes are involved (see particularly \cite{kn:newemp}), \emph{but} that they are almost certainly extremely limited in both spatial and temporal extent. This means that the naked singularity arising from a bifurcation is most unlikely to be the dominant feature of the bifurcated state. It does \emph{not}, however, mean that the naked singularity can be ignored completely.

We suggest, then, that the fact that the WGC implies the existence of naked singularities should not be regarded as an objection to it. However, there are other potential problems implicit in this discussion.

A rather subtle aspect of the above interpretation of the Theorem (which we have taken from \cite{kn:wald}) is its mention of the \emph{passing of time}: the interpretation only makes sense if we know which direction in time is the future.

That direction is of course the direction in which the \emph{entropy} of the system increases, in accord with the second law of thermodynamics \cite{kn:lebo}. The correct statement, then, is that the Bifurcation Theorem can be interpreted in the above manner (as implying that bifurcations give rise to naked singularities) \emph{provided} that the (proposed) final state has a \emph{greater total entropy than the initial black hole}\footnote{In a recent important development, the opposite process, the \emph{merger} of astrophysical black holes, has been studied \cite{kn:holley} using black hole entropy. Such mergers are a subject of intense observational interest, and the entropy technique promises to have concrete applications.}. In the case in which the emitted object is a black hole, that state consists of a pair of objects with well-defined entropies \cite{kn:wall}, together with the object into which the naked singularity has evolved.

The question now is this: is it in fact true that the bifurcated state has a higher entropy than that of the initial black hole?

To answer this, one needs to consider the quantity of entropy one should assign to a singularity. The precise value is unknown; a full quantum-gravitational treatment would be required to determine it. One can argue, however, that it is strictly non-zero except in the special case of the cosmological initial singularity \cite{kn:lebo,kn:roger}, and it might well be very substantial. Thus we arrive at a rather surprising conclusion: the naked singularity required by the Bifurcation Theorem is actually welcome from this point of view, since \emph{it could help to render the WGC compatible with the second law}. In some cases there are also other effects which likewise contribute to the final total entropy, as we shall discuss.

Unfortunately this does not settle the question, since it can happen that the total entropy of the two black holes produced by the bifurcation is much smaller than that of the original black hole, and it is not obvious that the ``other effects'' mentioned above can compensate for that. This is the first question we will address.

The second question is this: \emph{why} must an extremal or near-extremal black hole emit particles ---$\,$ or anything else?

In the case where the emitted object is a black hole, let us suppose that the bifurcated state does actually have a higher entropy than the original state. In general, if a system can evolve to state with a higher entropy without violating constraints or conservation laws, then it can be expected to do so eventually. But what is the \emph{mechanism} which enables the transition ---$\,$ what physical effect causes near-extremal black holes to be unstable? Unless we can identify this mechanism, we cannot claim to have understood the WGC.

Note carefully that it would not be satisfactory to have a mechanism that only works for black holes which are \emph{exactly} extremal: near-extremality is far more likely to be physical than its exact counterpart \cite{kn:naresh}. For example, Hawking radiation can cause a charged black hole (in any number of spacetime dimensions) to approach extremality very closely, but not to reach it \cite{kn:haoong}. In short, whatever mechanism allegedly causes the instability must apply also to black holes which are (sufficiently) \emph{close} to being extremal. (In fact, throughout this work, the reader should interpret ``extremal'' as meaning ``not necessarily exactly extremal, but as close to extremality as desired''.)

We will consider two examples. In the first (asymptotically flat five-dimensional Reissner-Nordstr\"{o}m), we have not succeeded in resolving either of these questions; in the second (asymptotically AdS$_5$ Reissner-Nordstr\"{o}m black holes with \emph{planar} event horizons), we claim to be able to answer the first question, and partially to answer the second.

Let us discuss first the case of a five-dimensional asymptotically flat extremal Reissner-Nordstr\"{o}m black hole. (Throughout this work, all black holes are five-dimensional, since eventually we hope to gain some insight using the AdS/CFT correspondence \cite{kn:nat}, which is best understood for asymptotically AdS$_5$ spacetimes.)

\addtocounter{section}{1}
\section* {\large{\textsf{2. Asymptotically Flat Reissner-Nordstr\"om Black Holes}}}
The asymptotically flat five-dimensional Reissner-Nordstr\"om black hole metric is
\begin{flalign}\label{A}
g(\m{AFRN_5})\;=\;&-\,\left(1\,-\,{8M\ell_5^3\over 3\pi r^2}\,+\,{k_5Q^2\ell_5^3\over 3\pi^3 r^4}\right)\m{d}t^2\,+{\m{d}r^2\over 1\,-\,{8M\ell_5^3\over 3\pi r^2}\,+\,{k_5Q^2\ell_5^3\over 3\pi^3 r^4}}\\ \notag \,\,\,\,&\,+\,r^2\left(\m{d}\theta^2 \,+\, \sin^2\theta\,\m{d}\phi^2\,+\,\cos^2\theta\,\m{d}\psi^2\right).
\end{flalign}
Here $\ell_5$ is the five-dimensional Planck length, $k_5$ is the five-dimensional Coulomb constant\footnote{We use units in which electric charge and entropy are dimensionless, while mass and energy have units of inverse length. Note that the Coulomb constant $k_5$ in five dimensions is then not dimensionless like its four-dimensional counterpart: it has units of length, and must \emph{not} be ignored (``set equal to unity'').}, $r$ and $t$ are as usual, $M$ and $Q$ are the mass and charge respectively, and the $t = $ constant, $r = $ constant sections are three-spheres described by angular Hopf coordinates $\theta, \phi, \psi.$

The value of the radial coordinate at the event horizon of such a black hole, $r_{\textsf{H}}$, is easily found by solving a quadratic (in $r_{\textsf{H}}^2$),
\begin{equation}\label{B}
1\,-\,{8M\ell_5^3\over 3\pi r_{\textsf{H}}^2}\,+\,{k_5Q^2\ell_5^3\over 3\pi^3 r_{\textsf{H}}^4}\;=\;0,
\end{equation}
and then one finds as usual, by requiring that the discriminant should not be negative, that the ratio $M\ell_5/Q$ is bounded below by a fixed constant:
\begin{equation}\label{C}
{M\ell_5\over Q} \;\geq\; {1\over 4}\,\sqrt{{3k_5\over \pi \ell_5}}.
\end{equation}
Thus, in the extremal case, $M\ell_5$ is a fixed dimensionless multiple of the charge, while $r_{\textsf{H}}$ in this case is a multiple of the square root of either. Consequently the entropy of a five-dimensional extremal asymptotically flat Reissner-Nordstr\"{o}m black hole is given by
\begin{equation}\label{D}
S_{\textsf{AFRN}_5}^{\textsf{ext}}\;=\;{1\over 6}\,\left[{3k_5^3\over \pi \ell_5^3}\right]^{{1\over 4}}Q^{{3\over 2}}.
\end{equation}

If we consider a pair of extremal five-dimensional Reissner-Nordstr\"{o}m black holes with charges $Q_1$ and $Q_2$, such that $Q_1\;+\;Q_2 \;=\; Q$, then using the elementary inequality $(x + y)^a > x^a + y^a,$ which is strict for all positive real $x, y$ and for all real $a > 1$ (though not, of course, for $a = 1$), we see at once that, in the first approximation, the bifurcation of such a black hole into two smaller (classical) black holes of the same kind \emph{would cause the total entropy to decrease}.

However, it can be argued that this computation underestimates the entropy of the final system. Issues related to this statement have been discussed very extensively in the literature, and there are many subtleties connected with it: we refer the reader to \cite{kn:gary} and works citing it for an in-depth discussion. Here we wish to focus on the following observations, which are very general and directly relevant to our later discussion.

First, for a non-extremal black hole, one would expect a bifurcation to give rise to Joule heating (and possibly some viscous dissipation) as the charge is rearranged: see \cite{kn:empa} and its references. Whether this necessarily happens in the extremal case, for which the temperature is \emph{zero}, is however much less clear. We will be cautious and assume that it does not: that is, we assume that extremal black holes emit smaller but still extremal black holes. \emph{However}, if this were incorrect, then the temperatures of the final pair of black holes would rise, and therefore (since extremal black holes have the smallest entropy among black holes of given mass) the entropies would be \emph{larger} than the above calculation suggests. That is, if our assumption were incorrect, we would be systematically underestimating the total final entropy, and of course that would only reinforce any argument to the effect that the second law is satisfied.

Secondly, a simple calculation \cite{kn:kats} shows that, if an extremal black hole of this kind emits a smaller black hole, and if the survivor continues to satisfy classical Cosmic Censorship, then the mass $m$ and charge $q$ of the emitted black hole must satisfy\footnote{If we interpret $m\ell_5/q$ as a measure of the relative strengths of the gravitational and gauge ``charges'', then the fact that this ratio is bounded above is one way of expressing the claim that gravity is ``weak''.}
\begin{equation}\label{DD}
{m\ell_5\over q} \;<\; {1\over 4}\,\sqrt{{3k_5\over \pi \ell_5}}.
\end{equation}
But, of course, no values of $m$ and $q$ can satisfy both (\ref{C}) and (\ref{DD}).

One might argue that, since the Bifurcation Theorem already implies the existence of a naked singularity here, the possibility that the emitted object might be naked does not make the situation worse. However, the argument in \cite{kn:roberto} is that naked singularities may exist \emph{briefly} in the highly dynamic spacetime in the immediate vicinity of the bifurcation event. They should \emph{not} be present in the regime in which the emitted object has settled down to its final state, and the geometry is not strongly time-dependent. Thus we do appear to have a contradiction here.

The contradiction is however derived by using classical Censorship (twice: for both the emitted object and the survivor). Following \cite{kn:kats,kn:NAH}, we take this to mean that the emitted black hole is governed by a quantum-corrected version of Censorship. This might permit lower values of the mass than (\ref{C}) allows, resolving the contradiction.

As is well known, it has recently been argued \cite{kn:rem1,kn:goon} that such corrections, in the form of higher-derivative corrections, are associated with an \emph{upward} modification of the classical entropy\footnote{See however \cite{kn:cano1,kn:cano2,kn:cano3,kn:peak}: this statement may need some refinement.}. That is, when these corrections are taken into account, the emitted black hole has a higher entropy than an extremal classical black hole of the same charge. Granted that this is so, then the above classical calculation of the entropy of the whole system definitely leads to an underestimate of the actual final total entropy.

The problem, however, is that the Wilson coefficients in the higher-derivative expansion are not known numerically.  Therefore, we cannot assert definitely that the upward revision of the entropy of the emitted black hole suffices to over-compensate for the classical decrease in the entropy \emph{of the entire system}. In short, we have two effects here pushing the total entropy in two different directions, and we cannot yet say definitely which effect dominates (in this case).
 
Finally, as we have seen, a naked singularity is necessarily produced by the bifurcation, and it (or, still more, the system into which it evolves) contributes some quantity of entropy to the final system. It is not known how to evaluate the entropy of a naked singularity ---$\,$ of course, the area formula cannot be used ---$\,$ but general statistical mechanical principles allow us to say that the most probable state of any isolated system is the one of maximal entropy. One therefore expects whatever emerges from a naked singularity to be in a state of high entropy (see the classic discussion of this idea, in the cosmological context, by Penrose \cite{kn:roger}). Thus, again, the above classical estimate of the total final entropy is an underestimate.

In summary, despite the decrease in the entropy in the first approximation, it is certainly entirely possible that these additional effects imply that, overall, the entropy increases, meaning that bifurcation can occur in accordance with the second law of thermodynamics. While there is good circumstantial evidence \cite{kn:hod} that this is indeed so (see however \cite{kn:heid}), it is very difficult to estimate the actual numerical values of these additional contributions to the final entropy. Thus the problem is not fully understood.

To understand this in quantitative detail, it would be extremely helpful to have an explicit account of the precise mechanism causing the bifurcation (or even the emission of particles by extremal asymptotically flat black holes). Unfortunately, no such mechanism is known in this case in sufficient detail.

We will now proceed to argue that the situation in the asymptotically AdS$_5$ case is dramatically clearer: bifurcation is in many cases (see for example \cite{kn:105}) demonstrably compatible with the second law of thermodynamics, and furthermore in one case ---$\,$ charged black holes such that the transverse spatial sections (that is, the t = constant, r = constant submanifolds outside the event horizon) are \emph{planar} ---$\,$ there is also a very natural explicit mechanism which explains the instability of near-extremal black holes.

\addtocounter{section}{1}
\section* {\large{\textsf{3. AdS$_5$-Reissner-Nordstr\"om Black Holes With Toroidal/Planar Transverse Spatial Sections}}}
Asymptotically AdS black holes need not have topologically spherical transverse spatial sections \cite{kn:lemmo}: in particular, the geometry can be that of a flat torus, or, as a limiting case, that of a plane\footnote{The ``plane'' is of course three-dimensional in our case, so we speak of volumes rather than areas in connection with it.}. The planar case is in fact the most important one, since it is the one almost always used in applications of the AdS/CFT correspondence \cite{kn:nat}, the conformal boundary having the conformal geometry of Minkowski spacetime. Nevertheless, the toroidal case is mathematically simpler, so we begin with it.

The locally asymptotically AdS$_5$-Reissner-Nordstr\"om metric in the case with flat toroidal transverse spatial sections is given by
\begin{flalign}\label{E}
g(\m{AdSRN_5^0})\;=\;&-\,\left({r^2\over L^2}\,-\,{16\pi M^*\ell_5^3\over 3r^2}\,+\,{4\pi k_5 Q^{*2}\ell_5^3\over 3r^4}\right)\m{d}t^2\,+{\m{d}r^2\over {r^2\over L^2}\,-\,{16\pi M^*\ell_5^3\over 3r^2}\,+\,{4\pi k_5 Q^{*2}\ell_5^3\over 3r^4}}\\ \notag \,\,\,\,&\,+\,r^2\left(\m{d}\theta^2 \,+\,\m{d}\phi^2\,+\,\m{d}\psi^2\right).
\end{flalign}
Here $k_5$ is the five-dimensional Coulomb constant as above, $L$ is the AdS curvature length scale, $\ell_5$ is the gravitational length scale for AdS$_5$, and the coordinates $(\theta, \phi, \psi)$ are each compact, that is, circular; for simplicity we take it that the periodicity is $2\pi K$ for all three, where $K$ is the compactification parameter\footnote{When $M^* = Q^* = 0$, the geometry of the spacetime is that of AdS$_5$, with a compactification of the spatial sections in a specific foliation. From a geometric point of view, then, $K$ is a property of the AdS$_5$ ``background'', not (only) of some specific black hole. This means that all black holes in a given background are described by the same value of $K$.}. Thus, in the toroidal case, the three-dimensional volume of any transverse spatial section with a given value of the coordinate $r$ is given by $Wr^3$, where $W \equiv \left(2\pi K\right)^3$.

The mass and charge parameters $M^*$ and $Q^*$ are related in the toroidal case to the physical mass $M$ and the physical charge $Q$ by $M^* = M/W, \; Q^* = Q/W.$

Let $r_{\textsf{H}}$ denote the value of the radial coordinate at the outer horizon: that is, $r_{\textsf{H}}$ is the function of $M$, $Q$, $K$, $L$, and $\ell_5$ given by the largest solution of the equation
\begin{equation}\label{F}
{r_{\textsf{H}}^2\over L^2}\,-\,{16\pi M^*\ell_5^3\over 3r_{\textsf{H}}^2}\,+\,{4\pi k_5 Q^{*2}\ell_5^3\over 3r_{\textsf{H}}^4} \;=\; 0.
\end{equation}
Then $2\pi K r_{\textsf{H}}$ is the length of the event horizon along any of the angular axes; setting this expression equal to that length, one can solve for $K$ in terms of $M$, $Q$, $L$, and $\ell_5$; thus, $K$ has a definite geometric meaning, and is determined, in principle, by a measurable feature of the spacetime.

The entropy of the black hole is a multiple of $Wr_{\textsf{H}}^3$ (see below). Of course, toroidal black holes, like their spherical counterparts, can have arbitrarily large or small entropies; the only difference is that, in the toroidal case, the entropy is not dictated by $M$, $Q$, $L$, and $\ell_5$ alone, but has to be regarded as an independent parameter. One can say that $K$ plays this role, in the sense that it is determined by the entropy if $M$, $Q$, $L$, and $\ell_5$ are fixed. (That is, under these circumstances one can solve for $K$ (using (\ref{F})) if $Wr_{\textsf{H}}^3$ is given.) This supplies the physical interpretation of $K$: it is the ``\emph{entropy parameter}''. (In a sense this is analogous to the well-known fact that the Hawking temperature of the black hole is determined by the periodicity of Euclidean time.)

In the planar case, obtained in the limit $K \rightarrow \infty$, the black hole has formally infinite mass and charge. But $M^*$ and $Q^*$ are still well-defined, because we can allow $M$ and $Q$ to tend to infinity along with $W$, in such a manner that $M^*$ and $Q^*$ remain finite.

Various combinations of $M^*$ and $Q^*$ (but never $M$ or $Q$ or $W$ by themselves) appear in the holographic ``dictionary'' for these black holes, and so this dictionary applies equally in the toroidal and planar cases. Briefly (see \cite{kn:106} for a detailed discussion), the energy (or enthalpy) density of the boundary matter is dual to the mass per unit horizon volume of the bulk black hole:
\begin{equation}\label{FF}
\rho\;=\; {M^*\over r_{\textsf{H}}^3}.
\end{equation}
Likewise the analogue of the quark chemical potential in the boundary field theory is given by duality as
\begin{equation}\label{FFF}
\mu\;=\;{k_5Q^*\over 2 r_{\textsf{H}}^2},
\end{equation}
and, as is well known, the temperature of the boundary theory is the Hawking temperature of the black hole,
\begin{equation}\label{FFFF}
T\;=\;{1\over \pi}\left({r_{\textsf{H}}\over L^2}\;-\;{2\pi k_5Q^{* 2}\ell_5^3 \over 3 r_{\textsf{H}}^5}\right).
\end{equation}
Finally, the ratio\footnote{Note that, in phenomenological practice, the entropy and energy of strongly coupled fluids (such as the quark-gluon plasma) must be quantified through these densities.} of the energy or enthalpy density $\rho$ of the boundary field theory to its entropy density $\rho_S$ is given holographically by
\begin{equation}\label{FFFFF}
{\rho\over \rho_S}\;=\;{4M^*\ell_5^3\over r_{\textsf{H}}^3}.
\end{equation}
As claimed, all of these quantities depend \emph{only} on $M^*$ and $Q^*,$ and so holography works equally well in the planar and toroidal cases. We note in passing that the right sides of (\ref{FF}), (\ref{FFF}), (\ref{FFFF}), and (\ref{FFFFF}) all depend non-trivially on $K$, through $M^*$, $Q^*$, and $r_{\textsf{H}}$.

It is worth stressing that toroidal/planar black holes are indeed very different from their spherical relatives; for example, it does not make sense to take the limit $L \rightarrow \infty$ in equation (\ref{E}). Another difference derives from the very fact that the transverse spatial sections are \emph{flat}. Since the Einstein equations are local, not global, restrictions on the geometry, one obtains a solution (even for given values of $M^*$ and $Q^*$) for \emph{any} flat three-dimensional manifold, and there are \emph{many} such manifolds: see \cite{kn:conway}. (In the locally spherical case, too, there is an ambiguity of this sort, but it is very much less extensive, in the sense that the family of 3-dimensional spherical manifolds of given curvature differ from each other in a purely discrete manner.) Here, for simplicity, we are reducing the vast family of flat candidates (for fixed values of the mass and charge) to the relatively small family, parametrised by $K$, which differ only by their entropies.

We now come to an elementary but crucial observation: whereas equation (\ref{B}), which determines $r_{\textsf{H}}$ in the asymptotically flat case, is essentially a quadratic, equation (\ref{F}) is by contrast a \emph{cubic} (in $r_{\textsf{H}}^2$). Such an equation also has a discriminant \cite{kn:disc}, but its structure is quite different to that of the discriminant of a quadratic equation.

In the case of asymptotically AdS$_5$ black holes with \emph{spherical} event horizons, it is also true that $r_{\textsf{H}}^2$ is found by solving a cubic; see \cite{kn:102}. But in that case the condition obtained by requiring the cubic discriminant to be non-negative (that is, the condition for Cosmic Censorship to hold) cannot differ too greatly from the one that governs the asymptotically flat black holes discussed in the previous Section, in the sense that the two must coincide in the limit $L \rightarrow \infty.$

That is not so in the case we are considering here, and the upshot \cite{kn:106} is that Cosmic Censorship takes a very unusual form for asymptotically AdS$_5$ black holes with toroidal or planar transverse spatial sections:
\begin{equation}\label{G}
{M^*\ell_5\over Q^*}\;\geq \;{3\over 16}\left({12\,k_5^2\over \pi L^2}\right)^{{1\over 3}}\,Q^{*{1\over 3}}.
\end{equation}
This differs crucially from the analogous inequality (see (\ref{C})) in the preceding Section: in particular, in the extremal case $M^*$ is \emph{not} a fixed multiple of $Q^*,$
but rather of $Q^{*{4\over 3}}$.

Now let us focus on the case in which the emitted object is itself a ``small'' black hole: let us assume that it has mass parameter $m^* < M^*$ and charge parameter $q^* < Q^*.$ Then it is possible to show \cite{kn:106} that the analogue of the inequality (\ref{DD}) is

\begin{equation}\label{H}
{m^*\ell_5\over q^*}\; < \;{1\over 4}\left({12\,k_5^2\over \pi L^2}\right)^{{1\over 3}}\,Q^{*{1\over 3}}.
\end{equation}

There are several important observations to be made here. First, note that the constant coefficient on the right side of (\ref{DD}) involves $k_5/\ell_5,$ but here the corresponding quantity is $k_5/L.$ (We assume that $k_5$ and $\ell_5$ take the same values in both cases.) In discussions of the AdS/CFT correspondence, it is normally assumed that $L$ is by far the largest length scale of the system, and that $\ell_5$ is very much the smallest \cite{kn:nat}. In that context, then, the coefficient in (\ref{H}) is the smaller of the two, and so, if $Q^*$ is not very large, the gravitational ``charge'' is as usual dominated by the gauge charge.

On the other hand, we can of course make the right side of (\ref{H}) as large as we please, by taking $Q^*$ sufficiently large. In that case, (\ref{H}) permits (but does not of course demand) large values for the left side; and so it may not be the case that the gravitational ``charge'' (essentially $m^*\ell_5$) is \emph{always} dominated by the gauge field charge (effectively $q^*$) for these black holes. This shows that the concept of ``weak'' gravity is more subtle in the asymptotically AdS case than has been suspected; there may for example be a relation to the debates regarding the status of the ``Repulsive Force Conjecture''. (See \cite{kn:repulsive} for the state of this conjecture.)

Secondly, and more importantly for our concerns here, let us suppose for a moment that the emitted black hole satisfies classical Censorship. We have, from (\ref{G}),
\begin{equation}\label{I}
{m^*\ell_5\over q^*}\;\geq \;{3\over 16}\left({12\,k_5^2\over \pi L^2}\right)^{{1\over 3}}\,q^{*{1\over 3}}.
\end{equation}

In dramatic contrast to the asymptotically flat case, \emph{it is perfectly possible to satisfy both (\ref{H}) and (\ref{I}) simultaneously.} For it is clear that the right side of (\ref{H}) is always strictly larger than the right side of (\ref{I}): so, given $Q^*$ and $q^*$, one need only choose $m^*$ so that $m^*\ell_5/q^*$ lies in the intervening interval.

The case in which the emitted black hole is itself extremal ---$\,$ recall from the preceding Section that this is our default assumption ---$\,$ is particularly clear. For then we have
\begin{equation}\label{J}
{M^*\ell_5\over Q^*}\;= \;{3\over 16}\left({12\,k_5^2\over \pi L^2}\right)^{{1\over 3}}\,Q^{*{1\over 3}}.
\end{equation}
and
\begin{equation}\label{K}
{m^*\ell_5\over q^*}\;=  \;{3\over 16}\left({12\,k_5^2\over \pi L^2}\right)^{{1\over 3}}\,q^{*{1\over 3}}.
\end{equation}
Suppose that $Q^*$ and $q^* < Q^*$ are given, so that $M^*$ and $m^*$ are computed from (\ref{J}) and (\ref{K}). Then it is immediately clear that $m^* < M^*$, and the fact that $3/16 < 1/4$ means that (\ref{H}) is satisfied. (Indeed it would be satisfied even if $q^*$ replaced $Q^*$.)

In short, the instability of extremal black holes of this kind does not necessarily lead to a conflict with (classical) Censorship in the final state (that is, beyond the temporary violation required by the Bifurcation Theorem), as it does for asymptotically flat black holes, or for asymptotically AdS black holes with spherical event horizons. So, for the moment, we can assume that the emitted black hole is a classical object. (Of course, the emitted black hole \emph{can} be non-classical: the point is just that it is not required to be so.)

This is very useful, because it means that we can exactly compute all of the entropies in this specific case and so determine whether there is any conflict with the second law of thermodynamics.

The value of the radial coordinate at the event horizon in the extremal case, $r_{\textsf{H}}^{\textsf{ext}},$ is found by solving (\ref{F}) (or more simply by setting the Hawking temperature equal to zero):
\begin{equation}\label{L}
r_{\textsf{H}}^{\textsf{ext}}\;=\;\left({2\pi k_5\over 3}\right)^{1/6}\,\sqrt{\ell_5}\,L^{{1\over 3}}Q^{*{1\over 3}}.
\end{equation}

In general, the entropy of a black hole with metric given by (\ref{E}) is, in the toroidal case,
\begin{equation}\label{M}
{S_{\m{AdSRN_5^0}}\over W}\;=\;{r_{\textsf{H}}^3 \over 4 \ell_5^3}.
\end{equation}
(Recall that $W$ is defined as $(2\pi K)^3,$ where $K$ is the compactification parameter for the toroidal transverse spatial sections. We write the expression in this way so as to include the planar case, where the entropy is formally infinite: in that case, with $K \rightarrow \infty$, both the numerator and the denominator on the left diverge, but their ratio remains finite.)

In the extremal case, we have, using (\ref{L}),
\begin{equation}\label{N}
{S_{\m{AdSRN_5^0}}^{\textsf{ext}}\over W}\;=\;{1\over 2}\,\sqrt{{\pi k_5L^2\over 6\ell_5^3}}\,Q^*.
\end{equation}

This is the classical entropy in the extremal case. In the previous Section, we could not use the analogous formula (equation (\ref{D})) because we knew that the emitted object could not be fully classical. In the toroidal/planar case, however, we have seen that the bifurcation does not necessarily lead to a violation of classical Censorship in the final state (though it can). It follows that, if we focus on cases where classical Censorship is not violated in the final state, we can justify using (\ref{N}) for any extremal toroidal/planar black hole.

Now, once again, the contrast with the asymptotically flat case is remarkable: instead of the $3/2$ power of the charge seen in (\ref{D}), we have in (\ref{N}) a simple \emph{linear} relation between the entropy and the charge parameter $Q^*$ (which, like the physical charge itself, is a conserved quantity). This immediately means that, if the extremal black hole splits into two extremal black holes, the bifurcation in the classical case does \emph{not} cause the entropy to decrease: \emph{the total entropy of the black holes themselves neither increases nor decreases}.

The Bifurcation Theorem, however, assures us that a naked singularity is generated by the splitting; as explained earlier, we assume that it has some entropy, and therefore the system into which it evolves (in accordance with \cite{kn:roberto}) has still more. It follows that, when the entropy contributed in this manner is included, the final state definitely has a higher total entropy than the original black hole. Thus we have a proof that \emph{the bifurcation definitely increases the total entropy of the system in the special case in which the emitted object is classical}.

If on the other hand the bifurcation generates a quantum-gravitational object, then the results of \cite{kn:rem1,kn:goon} (which do not depend on the topology of the event horizon) suggest that the total entropy of the final pair of black holes exceeds the classical total for the pair. But, as we have just seen, that classical total \emph{already} attains the entropy of the original extremal black hole. Thus we have a proof that, in \emph{all} cases, the bifurcation of an extremal toroidal/planar black hole increases the total entropy of the system\footnote{This argument means that, in practice, the emitted black hole will \emph{not} be classical, since the state with higher possible entropy will be favoured. If on the other hand quantum-gravitational effects had reduced the entropy, then this would not have invalidated our argument ---$\,$ it would merely mean that the emitted object would probably be classical. In such a scenario, the naked singularity would play the decisive role in ensuring that the second law is satisfied.}.

This certainly greatly strengthens the claim that bifurcation can occur for extremal toroidal/planar black holes. But we still do not understand the fundamental reason for this instability\footnote{It is worth noting that planar black holes do not have a Hawking-Page transition. See \cite{kn:surya}.}. There is some reason to hope that this too can be remedied in this specific case, as we now argue.

\addtocounter{section}{1}
\section* {\large{\textsf{4. Why Near-Extremal AdS$_5$-Reissner-Nordstr\"om Toroidal/Planar Black Holes Are Unstable}}}
We now proceed to propose an explicit (though admittedly incomplete) mechanism which might underlie the instability of near-extremal toroidal/planar black holes in the asymptotically AdS$_5$ case. We stress that this proposal is \emph{in no way} based on the arguments in the preceding Section; it stands or falls on its own merits.

From this point onwards, we will take it that the object emitted by a toroidal/planar black hole, when (as the WGC predicts) it decays, is a brane (and not a black hole). We also assume that this brane carries no electric charge, so it does not couple to the black hole itself. The brane is to be regarded as a probe, that is, we neglect its back-reaction on the background fields in the discussion. (This is appropriate for the early stages of any instability, which are our concern here. It might not be appropriate to the study of the final state to which the instability leads.)

Suppose for the sake of argument that a near-extremal toroidal/planar black hole does in fact emit a probe brane. The structure of such objects has of course been studied extensively \cite{kn:clifford}. We would like to understand the further evolution of the emitted object, in the string theory (or AdS/CFT) context.

In that context, the brane is governed by an action consisting of two terms. (See \cite{kn:seiberg}; we follow Seiberg and Witten and work in Euclidean signature.) Consider any hypersurface $\Sigma$ in the (Euclidean) AdS$_5$-Reissner-Nordstr\"om bulk homologous to the conformal boundary, representing a brane with area $A(\Sigma)$ containing a volume $V(\Sigma)$. The first term in the action is the obvious one: the brane tension contributes a positive term, proportional to $A(\Sigma)$. But Seiberg and Witten point out that there is another contribution, coming from the coupling of the brane to the five-form flux which is always present here; and this term is a \emph{negative} multiple of $V(\Sigma)$. The action $\mathfrak{S}$ for a BPS brane is then given \cite{kn:seiberg} by
\begin{equation}\label{O}
\mathfrak{S}\;\propto \; A(\Sigma)\;- \;{4 \over L}\,V(\Sigma).
\end{equation}
Thus there is a danger that \emph{the brane action could be negative}, as a direct result of the coupling to the five-form. This would mean that by creating a pair consisting of a brane and an antibrane and then allowing one of them to propagate towards infinity, we can decrease the action. In other words, under these conditions the black brane will spontaneously emit a brane or an antibrane. (See \cite{kn:maldacena} for a detailed discussion of this; see \cite{kn:niko1,kn:niko2,kn:oscar} for some interesting recent applications involving AdS/CFT duality.) In this case, we have an explicit description of the \emph{mechanism} responsible for the emission of the brane: it is due to the coupling to the five-form.

In an asymptotically AdS (Euclidean) space, areas and volumes grow at much the same rate as one moves towards infinity, and in fact there is are leading-order cancellations in the expansion of the right side of (\ref{O}). Thus there is a close contest between the two terms in the action, and it is not clear which term dominates. Notice that this is a question about the physics far from the black hole, so that effects which die off at large distances do not affect the outcome. For the same reason, clearly this is a question well-suited to AdS/CFT techniques, and Seiberg and Witten show that it can be settled in most cases by studying the effective squared mass of a certain scalar field defined on the conformal boundary. This effective squared mass is just the scalar curvature of the boundary metric. This scalar curvature is positive in the case of AdS black holes with spherical transverse spatial sections, so the effective squared mass is positive and the system is stable. For the same reason, AdS black holes with negatively curved transverse spatial sections are unstable.

That method breaks down, however, when the effective mass of the scalar is zero, which is precisely what is happening here, because the boundary has \emph{zero} scalar curvature. In this case the contest between the positive and negative terms is particularly close, and the problem can only be settled by means of a detailed calculation \cite{kn:2009}.

The Euclidean version of the geometry described by equation (\ref{E}) is of course
\begin{flalign}\label{PP}
g(\m{AdSRN_5^0})\;=\;&\left({r^2\over L^2}\,-\,{16\pi M^*\ell_5^3\over 3r^2}\,+\,{4\pi k_5 Q^{*2}\ell_5^3\over 3r^4}\right)\m{d}t^2\,+{\m{d}r^2\over {r^2\over L^2}\,-\,{16\pi M^*\ell_5^3\over 3r^2}\,+\,{4\pi k_5 Q^{*2}\ell_5^3\over 3r^4}}\\ \notag \,\,\,\,&\,+\,r^2\left(\m{d}\theta^2 \,+\,\m{d}\phi^2\,+\,\m{d}\psi^2\right).
\end{flalign}
Here $t$ is to be interpreted as a periodic coordinate, with period related inversely to the Hawking temperature.

The two integrals in (\ref{O}) are in this case respectively 4- and 5-dimensional. However, the metric coefficients depend only on $r$, so four of the integrals (the ones over the four periodic coordinates, including Euclidean time) are trivial and merely contribute an overall constant factor, which can be absorbed into the constant of proportionality in the expression (equation (\ref{O})) for the action\footnote{In the planar case the areas and volumes are of course infinite; we regulate in the usual way, by working with the compactified (toroidal) case, and extrapolating to the planar case at the end. Similarly the integral over Euclidean time would diverge if we considered the exactly extremal case (which we do not).}.

Performing the integrals and Wick-rotating back to Lorentzian signature, we obtain straightforwardly
\begin{equation}\label{P}
\mathfrak{S}\left(\m{AdSRN_5^0}\right)\;\propto \;r^3\left[{r^2\over L^2}\,-\,{16\pi M^*\ell_5^3\over 3r^2}\,+\,{4\pi k_5 Q^{*2}\ell_5^3\over 3r^4}\right]^{{1\over 2}}\;- \;{r^4\,-\,r_{\textsf{H}}^4\over L}.
\end{equation}
This action vanishes at the event horizon, but then grows with $r$ to reach a certain positive maximum. It then decreases, however, and can be either positive or negative at large distances from the black hole. To understand this, we consider the asymptotic expansion:
\begin{equation}\label{Q}
\mathfrak{S}\left(\m{AdSRN_5^0}\right)\;\propto \;{r_{\textsf{H}}^4\over L}\;-\;{8\pi \ell_5^3 L M^*\over 3}\;+\;{2 \pi L k_5 \ell_5^3 Q^{*2}\over 3r^2}\;-\;{32\pi^2 \ell_5^6 L^3 M^{*2} \over 9r^4}\;+\;\mathcal{O}(1/r^6).
\end{equation}
Notice that the terms proportional to $r^4$ have cancelled.

Clearly the question as to whether this function ever becomes negative is determined by the sign of the constant term in this expansion. That sign will be negative, signalling the spontaneous emission of branes, provided that
\begin{equation}\label{R}
r_{\textsf{H}}\;<\;\left({8\pi \ell_5^3L^2M^*\over 3}\right)^{{1\over 4}}.
\end{equation}
Substituting this value of $r$ into equation (\ref{F}), we find that this means that $Q^*$ has to exceed a certain critical value, $Q^*_{\textsf{crit}}$, defined as follows:
\begin{equation}\label{S}
Q^*\;>\;Q^*_{\textsf{crit}} \;\equiv \; \left({32 \over 3}\right)^{{1\over 4}}\,\left[{\pi L^2 \ell_5^3M^{*3}\over k_5^2}\right]^{{1\over 4}}.
\end{equation}

On the other hand, the extremal value of $Q^*$ for given $M^*$ is, from the inequality (\ref{G}),
\begin{equation}\label{T}
Q^*_{\textsf{ext}} \;\equiv \; {4\sqrt{2}\over 3}\,\left[{\pi L^2 \ell_5^3M^{*3}\over k_5^2}\right]^{{1\over 4}}.
\end{equation}
It is a remarkable fact that this is \emph{slightly larger} than $Q^*_{\textsf{crit}}$ for fixed $M^*$: we have
\begin{equation}\label{U}
{Q^*_{\textsf{crit}} \over  Q^*_{\textsf{ext}}} \; = \; \left({27\over 32}\right)^{{1/4}}\;\approx \; 0.95841.
\end{equation}
Thus there is a narrow band of values for $Q^*$ between $Q^*_{\textsf{crit}}$ and $Q^*_{\textsf{ext}}$ such that the brane action is negative (on an infinite interval of $r$ values extending out to infinity). Black holes with charge parameters in that band will in fact decay spontaneously by inducing brane pair-production in the ambient spacetime.

Thus we conclude that near-extremal toroidal/planar black holes are indeed unstable, as the WGC requires. Furthermore, we now know \emph{why}: it is due to the coupling of probe branes to the background five-form, which, when the black hole is close to being extremal, triggers a pair-production instablity.

This discussion applies, strictly speaking, only to emitted branes, not to emitted black holes. Thus, while we have definitely established that near-extremal toroidal/planar black holes are unstable, we have not proved that this is the mechanism underlying the fission of these black holes ---$\,$ despite the fact that we know that such fissions must happen, since they are favoured thermodynamically. However, a sufficiently large collection of branes corresponds in the supergravity limit to a black brane, and the latter can indeed decay through brane nucleation: see \cite{kn:niko1,kn:niko2,kn:oscar} for discussions of this. From this point of view, the distinction between emitted branes and emitted black branes is not very great.

However, our probe branes in the above discussion are not electrically charged. In order to show that planar black holes can emit small charged planar black holes, as the WGC suggests, one would need to improve our argument above by allowing the probe branes to be charged, and modifying the brane action accordingly. This is a matter for future work.

\addtocounter{section}{1}
\section* {\large{\textsf{5. Conclusion}}}
Near-extremal black holes are of very considerable interest, not only for pure theory, but also in theoretical and even observational astrophysics. For example, studies of primordial black holes \cite{kn:siri} involve them, they may have a characteristic observational signature \cite{kn:garg}, and observations of Cygnus X-1 suggest \cite{kn:cygnus} that this system contains a black hole with dimensionless spin parameter possibly as high as $0.9985,$ where unity represents exact extremality. In all these cases, near-extremality is due to rapid rotation, but there are many similarities between that case and near-extremality due to high electromagnetic charges; furthermore it is conceivable that near-extremality due to (magnetic) charge could be observable in the future \cite{kn:juan1,kn:yang,kn:lang1,kn:ullah,kn:lang2}.

The Weak Gravity Conjecture is based on the idea that black holes are unstable if they are sufficiently near to extremality. In the asymptotically flat case it is difficult to understand precisely how this works; but we have shown that the situation is much simpler in the important alternative case of asymptotically AdS$_5$ black holes with flat transverse spatial sections. In this case it is possible to prove that the second law of thermodynamics does not obstruct the splitting of the black hole, and it is possible to give an explicit proposal as to precisely why one should expect these black holes to be unstable, once they are embedded in string theory.

Having such a description allows us to answer other questions we may have. For example, we can now specify precisely what ``near-extremal'' means in this case: it means that the charge parameter is at least approximately $96\%$ of its extremal value for a given mass parameter.

One often finds that solving a problem in the AdS context does not provide a \emph{direct} route to a solution in the asymptotically flat case, and so it is here. In the asymptotically flat situation, we still do not know how to provide a quantitative estimate of the entropy of the post-bifurcation state, and the instability mechanism we discussed in the preceding Section does not work there (in fact it does not even work for spherical AdS Reissner-Nordstr\"{o}m black holes, because the scalar curvature at infinity is then positive: see \cite{kn:seiberg}). Nevertheless the AdS perspective has often been useful, and one can hope that it will be so in this case also.

\addtocounter{section}{1}
\section*{\large{\textsf{Acknowledgement}}}
The author is grateful to Dr. Soon Wanmei for useful discussions.

\end{document}